\documentclass[11pt]{article}
\usepackage{jheppub}
\usepackage{amsfonts,ulem,bbold}
\usepackage{amsmath,amssymb, braket}
\newcommand{\gmn}{g_{\mu\nu}}
\newcommand{\bgmn}{\bar g_{\mu\nu}}
\newcommand{\bGmn}{\bar G_{\mu\nu}}
\newcommand{\beqn}{\begin{eqnarray}}
\newcommand{\eeqn}{\end{eqnarray}}
\newcommand{\be}{\begin{equation}}
\newcommand{\ee}{\end{equation}}
\newcommand{\nn}{\nonumber}

\newcommand{\p}{\partial}

\newcommand{\fmn}{f_{\mu\nu}}
\newcommand{\Gmn}{G_{\mu\nu}}
\newcommand{\Mmn}{M_{\mu\nu}}

\newcommand{\bfmn}{\bar f_{\mu\nu}}

\newcommand{\bg}{\bar{g}}
\newcommand{\tr}{\mathrm{Tr}}

\newcommand{\mfp}{m_{\mathrm{FP}}}
\newcommand{\tmfp}{\tilde{m}_{\mathrm{FP}}}
\newcommand{\td}{\mathrm{d}}

\def\ph{\phantom}
 
\title{Bimetric Theory and Partial Masslessness with Lanczos-Lovelock Terms in Arbitrary Dimensions}   
\author{S.F.~Hassan,}
\author{Angnis~Schmidt-May,}
\author{Mikael~von~Strauss}
\affiliation{Department of Physics \& 
        The Oskar Klein Centre,\\
        Stockholm University, AlbaNova University Centre, 
        SE-106 91 Stockholm, Sweden}
\emailAdd{fawad@fysik.su.se}
\emailAdd{angnis.schmidt-may@fysik.su.se}
\emailAdd{mvs@fysik.su.se}

\abstract{ Ghost-free bimetric theories describe nonlinear
  interactions of massive and massless spin-2 fields and, hence,
  provide a natural framework for investigating the phenomenon of
  partial masslessness for massive spin-2 fields at the nonlinear
  level. In this paper we analyze the spectrum of the ghost-free
  bimetric theory in arbitrary dimensions. Using a recently proposed
  construction, we identify the candidate nonlinear partially massless
  (PM) theories. It is shown that, in a 2-derivative setup, nonlinear
  PM theories can exist only in 3 and 4 dimensions. But on adding
  Lanczos-Lovelock terms to the bimetric action it is found that
  higher derivative nonlinear PM theories also exist in higher
  dimensions. This is consistent with existing results on the direct 
  construction of cubic vertices with PM gauge symmetry. We obtain
  the candidate nonlinear PM theories in 5, 6 and 8 dimensions but
  show that none exist in 7 dimensions.}

\toccontinuoustrue

\begin{document} 
\maketitle
\flushbottom

\section{Introduction}

To focus attention and fix conventions, we start with a review of
partial masslessness in linear Fierz-Pauli (FP) theory and the
question of its nonlinear extension. We then describe our approach to
the problem and summarize our results.

\subsection{Partial masslessness in Fierz-Pauli theory and beyond} 
\label{SeclPM}

The linear fluctuations $h_{\mu\nu}$ of a massive spin-2 field are
governed by the Fierz-Pauli equation
\cite{Pauli:1939xp,Fierz:1939ix} which in a $d$-dimensional spacetime
with a background metric $\bGmn$ becomes,   
\be
\label{linmass} 
\tilde{\mathcal{E}}_{\mu\nu}^{\rho\sigma}\,h_{\rho\sigma}
-\frac{2}{d-2}\tilde\Lambda\left(h_{\mu\nu}-\tfrac{1}{2}
\bar{G}_{\mu\nu}\bar{G}^{\rho\sigma}h_{\rho\sigma}\right) 
+\tfrac{\tilde m_{\mathrm{FP}}^2}{2}\,\left(h_{\mu\nu}-
\bar{G}_{\mu\nu}\bar{G}^{\rho\sigma}h_{\rho\sigma}\right)=0\,. 
\ee
$\tilde\Lambda$ is the cosmological constant and $\tilde m_{FP}$
is the mass. For $\tilde m_{FP}=0$, (\ref{linmass}) must reduce to the
linearized Einstein equation for a massless spin-2 field, hence
$\tilde{\mathcal{E}}$ is given by      
\be
\tilde{\mathcal{E}}^{\rho\sigma}_{\mu\nu}h_{\rho\sigma} 
=-\tfrac1{2}\Big[\delta^\rho_\mu\delta^\sigma_\nu\nabla^2
+\bar G^{\rho\sigma}\nabla_\mu\nabla_\nu -\delta^\rho_\mu
\nabla^\sigma\nabla_\nu-\delta^\rho_\nu\nabla^\sigma
\nabla_\mu-\bar{G}_{\mu\nu}\bar G^{\rho\sigma}\nabla^2 
+\bar G_{\mu\nu}\nabla^\rho\nabla^\sigma\Big]h_{\rho\sigma}\,.
\label{KO}
\ee
The structure of (\ref{linmass}) is determined by the requirement of
the absence of ghost. Generically, it allows for $d(d-1)/2-1$
propagating modes corresponding to a massive spin-2 field in
$d$~dimensions. The nature and dynamics of $\bGmn$ cannot be specified
further in this framework, but a nonlinear setup that embeds
(\ref{linmass}) must address this issue. 

If $\bGmn$ is a dS or, in general, an Einstein spacetime, then
on the Higuchi bound \cite{Higuchi:1986py, Higuchi:1989gz},  
\be 
\tilde m_\mathrm{FP}^2=\frac{2}{d-1}\,\tilde\Lambda \,,  
\label{HB}
\ee 
equation (\ref{linmass}) develops a gauge invariance 
$h_{\mu\nu}\longrightarrow h_{\mu\nu}+ \Delta h_{\mu\nu}$
with \cite{Deser:1983mm,Deser:2001pe,Deser:2001us,Deser:2001wx,
Deser:2001xr,Deser:2004ji},  
\be
\Delta h_{\mu\nu}\equiv\left(\nabla_{\mu}
\nabla_{\nu}+\frac{2\Lambda}{(d-1)(d-2)}\bar G_{\mu\nu}\right)\xi(x)\,.
\label{gaugesymh}
\ee 
This can be used to gauge away the helicity zero component of
$h_{\mu\nu}$, so in $d=4$ only the four polarizations $\pm 2,\pm 1$
survive. This is the linear ``partially massless'' (PM)
theory.\footnote{PM theories also arise in the more general context
  of higher-spin theories, see for example,
  \cite{Francia:2008hd,Joung:2012rv,Joung:2012hz}. But here we
  concentrate only on the spin-2 case.}       

The obvious question of course is if the gauge symmetry associated
with partial masslessness can be generalized from the FP theory in dS
spacetimes to a nonlinear theory of spin-2 fields. Several recent
studies have attempted to address this issue using different
approaches \cite{Zinoviev:2006im, Joung:2012rv, deRham:2012kf,
  Deser:2012qg}. In particular, the authors in  
\cite{Zinoviev:2006im, Joung:2012rv} directly construct cubic vertices for 
$h_{\mu\nu}$ with the above PM gauge invariance. This constructive
approach makes interesting predictions about the nonlinear PM
theory, showing that :
\begin{itemize}
\item Cubic $h_{\mu\nu}$ interactions with a PM gauge invariance
  (\ref{gaugesymh}) exist only in $3$ \cite{Joung:2012rv} and $4$
  \cite{Zinoviev:2006im, Joung:2012rv} 
  dimensions, as long as the theory involves no more than $2$
  derivatives. Hence, while linear PM theory exists in any dimension,
  $2$-derivative nonlinear PM theories can exist only in $3$ and $4$
  dimensions.
\item When higher derivative interactions are allowed, then PM gauge
  invariant cubic terms can be constructed even for dimensions $d>4$
  \cite{Joung:2012rv}. The structure of the higher derivative terms is
  such that for $d=4$ one again recovers the $2$-derivative
  theory. Hence the higher derivative terms are relevant only for
  $d>4$.
\end{itemize}
Any method of constructing nonlinear PM theories must give rise to
the above features implied by the explicit cubic vertex calculations.  

In \cite{Hassan:2012gz} we obtained a potential nonlinear PM theory as
a special ghost-free bimetric theory \cite{Hassan:2011zd} in $d=4$.
There it was not obvious that this construction was consistent with
the above dimension dependent features of PM theories. In the present
paper, we implement this construction in arbitrary dimensions
obtaining candidate PM theories including higher derivative terms. It
is found that the construction indeed meets the expectations from the
cubic vertex analysis. Let us consider this in some detail.

It is natural to look for nonlinear PM theories among the nonlinear
extensions of the FP theory (\ref{linmass}) that do not suffer from
the Boulware-Deser ghost instability \cite{Boulware:1972zf,
  Boulware:1973my}. For $\bGmn= \eta_{\mu\nu}$ such a nonlinear theory
was obtained and shown to be ghost-free in a certain ``decoupling
limit'' in \cite{deRham:2010ik, deRham:2010kj} (also see
\cite{deRham:2011rn}), whereas the complete nonlinear proof of absence
of ghost was given in \cite{Hassan:2011hr,Hassan:2011ea}. However,
these models do not admit dS solutions with massive FP perturbations
around them. A generalization with extra parameters that accommodates
a non-flat but non-dynamical $\bGmn$ was first considered in
\cite{Hassan:2011vm}.  In this case decoupling limit arguments do not
exist to argue the absence of the BD ghost, but unitarity was proven
directly in the nonlinear theory in a Hamiltonian analysis
\cite{Hassan:2011tf,Hassan:2011ea,Hassan:2012qv}. Finally, the
completely dynamical ghost-free theory was given in
\cite{Hassan:2011zd}, as a bimetric theory conveniently formulated in
terms of two interacting spin-2 fields $\gmn$ and $\fmn$. This theory
contains the two constraints required to eliminate the BD ghost
\cite{Hassan:2011zd,Hassan:2011ea}. For related work see
\cite{Alberte:2010qb, Alberte:2010it, Chamseddine:2011mu,
  Folkerts:2011ev, Kluson:2011qe, Comelli:2011wq, Volkov:2011an,
  vonStrauss:2011mq, Comelli:2011zm, Kluson:2011rt, Comelli:2012db,
  Kluson:2012wf, Hinterbichler:2012cn, Hassan:2012wt, Berg:2012kn,
  Kluson:2012zz, Volkov:2012zb, Burrage:2012ja, Park:2012cq,
  Alexandrov:2012yv}

As such, neither of the two metrics in the bimetric theory corresponds
directly to the massive fluctuation in (\ref{linmass}). But the theory
can be reformulated in terms of a massive spin-2 field $M^G_{\mu\nu}$
coupled to a massless metric $\Gmn$ such that linear perturbations
$\delta M^G_{\mu\nu}$ around a background ${\bar M}_{\mu\nu}^G=0$
always satisfy the FP equation (\ref{linmass}) where now one
identifies $h_{\mu\nu}=\delta M^G_{\mu\nu}$ \cite{HSV3}. Furthermore,
(\ref{linmass}) is also supplemented by a massless equation for
$\delta\Gmn$, while $\bGmn$ is determined as a solution to an ordinary
Einstein's equation. In the absence of matter couplings, a background
${\bar M}_{\mu\nu}^G=0$ forces $\bGmn$ to be an Einstein metric.\footnote{The theory has a much more complicated behaviour around
  backgrounds with ${\bar M}_{\mu\nu}^G\neq 0$ but this is not
  relevant for comparison to the FP equation.} Hence, the bimetric
theory provides a completely dynamical and ghost-free nonlinear
extension of (\ref{linmass}). For this reason it also provides the
natural arena for investigating partial masslessness beyond the FP
theory.

Bimetric theories contain several parameters that can be easily tuned
to put equation (\ref{linmass}) in the dS backgrounds on the Higuchi
bound (\ref{HB}). The linear theory will now have the gauge symmetry
(\ref{gaugesymh}), but only the $\xi=constant$ part of this
transformation preserves the dS backgrounds. In \cite{Hassan:2012gz}
it was argued that the consistency of these dS preserving
transformations with the dynamical nature of the backgrounds leads to 
a criterion for partial masslessness that is powerful enough to fix
most of the remaining parameters of the model. The criterion was
implemented in $d=4$ leading to a unique class of nonlinear candidate
PM theories \cite{Hassan:2012gz}. The complete gauge symmetry of the
nonlinear model is not yet known but some special cases can be
considered. Here we implement this criterion in arbitrary dimensions
to obtain candidate PM theories that are consistent with expectations
from the cubic vertex calculations described above. 

\subsection{Summary of results}

Before getting into technical details, let us briefly summarize
our results.
\begin{itemize}
\item The bimetric theory has a well defined mass spectrum around
  proportional backgrounds $\fmn=c^2\gmn$ where $c$ is generically
  determined by the equations of motion. We obtain the mass spectrum
  of the fluctuations in arbitrary dimensions with explicit
  expressions for the mass and the cosmological constant. We also
  construct the nonlinear extensions of the massless and massive
  fluctuations which are useful for the purpose of fixing the
  normalizations in the PM case. 
\item We then discuss the linear PM theory in the context of bimetric
  theory linearized around dS backgrounds. The consistency of a subset
  of (\ref{gaugesymh}) with the dynamical backgrounds leads to a
  simple criterion for identifying nonlinear PM theories as special
  bimetric theories with the parameters fixed such that the equations
  of motion leave the $c$ in the ansatz $\fmn=c^2\gmn$ undetermined.
  In the PM theory $c$ can be traded off with a constant gauge
  parameter. Applying this criterion to the $2$-derivative bimetric
  action we find that nonlinear PM theories can exist only in $d=3$
  and $d=4$. Although the full gauge transformation of nonlinear PM
  theories is not yet known, we show that the mass, cosmological
  constant, couplings and the background metric $\bGmn$ all become
  $c$-independent, and hence gauge invariant, for the PM parameter
  values. 
\item To explore the possibility of recovering nonlinear PM theories
  for $d>4$ with the help of higher derivative terms, we consider the
  bimetric theory with extra Lanczos-Lovelock terms for both metrics.\footnote{Such an 
extension has been considered earlier in a
    different context in \cite{Paulos:2012xe}.} The analysis of the spectrum
  around dS backgrounds can easily generalized to the LL terms using
  recent results of \cite{Sisman:2012rc} obtained for the LL extension of the
  Einstein-Hilbert gravity. We show that the construction outlined
  above can now be carried out leading to potential nonlinear PM
  theories even for $d>4$. In particular, we obtain such nonlinear
  theory for $d=5,6,8$ but show that no PM theory exists for $d=7$.
\item We also outline the structure of the bimetric action expressed
  in terms of the nonlinear extensions of the massless and the massive
  fields. This will be useful for comparison with the outcome of the
  direct cubic vertex construction, although such an explicit
  comparison has not been attempted here.
\end{itemize}

\section{Bimetric description of massless and massive spin-2 fields in
  $d$ dimensions} 

In this section we consider the ghost-free bimetric theory in
$d$ dimensions and describe in what sense it is a theory of a massive
and a massless spin-2 fields. To this end, we consider a class of
bimetric backgrounds, the proportional backgrounds, around which
linear perturbations decompose into well defined massless and massive
modes. These can then be promoted to nonlinear fields, generalizing
the notion of massless and massive spin-2 fields beyond the special
class of proportional backgrounds. The analysis generalizes the $d=4$ 
case considered in \cite{HSV3}.

\subsection{The bimetric theory in $d$ dimensions} 

The ghost-free bimetric action for two spin-2 field $\gmn$ and $\fmn$
\cite{Hassan:2011zd} can be easily generalized to arbitrary
dimensions,\footnote{In the paper we use the sign and curvature
  conventions of Wald \cite{Wald:1984rg} so that $R_{\mu\nu}=\p_\alpha
  \Gamma^\alpha_{\mu\nu}+\cdots$}   
\be
S_{gf}=\int\td^dx\left[m_g^{d-2}\sqrt{|g|}R(g)+m_f^{d-2}\sqrt{|f|}
R(f) -2m^d\sqrt{|g|}\,V(S;\beta_n)\right]\,,
\label{Sgf}
\ee
where $\sqrt{|g|}=\sqrt{|\det g|}$ and $S$ simply stands for the matrix
square-root,  
\be
S\equiv \sqrt{g^{-1}f}\,.
\label{Ssqrt}
\ee
The interaction potential is given by,
\be
V(S;\beta_n)=\sum_{n=0}^{d}\beta_n\,e_n(S)\,, 
\label{V}
\ee
where $e_n(S)$ are the elementary symmetric polynomials of eigenvalues
of $S$. They are expressible as polynomials of $\tr(S^k)$  
and can be iteratively constructed starting with $e_0(S)=1$ and using Newton's identities,  
\be
e_n(S)=-\frac{1}{n}\sum_{k=1}^{n}(-1)^k\,\tr(S^k)\,e_{n-k}(S)\,.
\ee
In particular, $e_d(S)=\det S$ and  $e_n(S)=0$ for $n>d$.

The $\beta_n$ in (\ref{V}) and the Planck masses $m_g$ and $m_f$ are 
the $d+3$ free parameters of the theory. The mass parameter $m$ is
degenerate with the overall scale of the $\beta_n$. The action
contains the cosmological terms $\beta_0\sqrt{|g|}$ and
$\beta_d\sqrt{|f|}$, but the actual cosmological constants are to be
read off from the equations of motion.

The absence of the Boulware-Deser ghost in
bimetric theory in $d=4$ was proved in
\cite{Hassan:2011zd,Hassan:2011ea}. It is straightforward to extend
the ghost analysis to arbitrary dimensions, especially using the
``deformed determinant'' representation for the potential
\cite{Hassan:2011vm}. Setting $\fmn$ to a non-dynamical flat metric
and tuning the $\beta_n$ to exclude a cosmological constant
contribution in a flat $\gmn$ background, one recovers the massive
gravity model of \cite{deRham:2010kj}.

The bimetric potential (\ref{V}) has the following useful symmetry
property under the interchange of $\gmn$ and $\fmn$ that sends
$S\rightarrow S^{-1}$ \cite{Hassan:2011zd}, 
\be 
\sqrt{|g|}\,V(S\,;\beta_n)= \sqrt{|f|}\,
V(S^{-1}\,;\beta_{d-n})\,.
\label{f-g}
\ee   
This allows us to obtain the $f$-sector equations from the $g$-sector
of the theory.

The sourceless equations of motion obtained on varying the action
(\ref{Sgf}) with respect to $g^{\mu\nu}$ and $f^{\mu\nu}$ are,
\be
\label{gf_eom}
R_{\mu\nu}(g)-\tfrac{1}{2}\gmn R(g)+\tfrac{m^d}{m_g^{d-2}}\,
V_{\mu\nu}^{g}= 0\,,\qquad
R_{\mu\nu}(f)-\tfrac{1}{2}\fmn R(f)+\tfrac{m^d}{m_f^{d-2}}\,
V_{\mu\nu}^{f}= 0\,.
\ee
The potential contributions have the form,
\begin{align}
V_{\mu\nu}^g=g_{\mu\lambda}\,V^{\lambda\,g}_{~\nu}(S)\,,\qquad
V_{\mu\nu}^f=f_{\mu\lambda}\,V^{\lambda\,f}_{~\nu}(S^{-1})\,,
\end{align}
and are explicitly given by (\ref{Ap-Vmng}) and (\ref{Ap-Vmnf}) in the
the appendix (where the details of the derivation are also provided).

\subsection{The proportional backgrounds} 

Let us now consider a class of solutions to the bimetric equations in
which the metrics are proportional to each other,  
\be
\bfmn= c^{2}\,\bgmn\,.
\label{fcg}
\ee
Such backgrounds have two important properties: 1) they are the
most general class of backgrounds around which the bimetric theory has
well defined massless and massive fluctuations with a Fierz-Pauli
mass term, 2) they coincide with classical solutions in general
relativity.\footnote{For another feature of such solutions, see
  \cite{Baccetti:2012re}} Indeed, for this ansatz, the bimetric
equations (\ref{gf_eom}) imply that $c$ is constant and then reduce to
two copies of cosmological Einstein's equation for $\bgmn$, 
\be
R_{\mu\nu}(\bar g)-\tfrac{1}{2}\bgmn R(\bar g)+\bgmn\Lambda_g=
0\,,\qquad
R_{\mu\nu}(\bar g)-\tfrac{1}{2}\bgmn R(\bar g)+\bgmn\Lambda_f=0\,.
\label{bg-g_eom}
\ee
The cosmological constants are given by (for details see the appendix),      
\be
\Lambda_g=\frac{m^d}{m_g^{d-2}}\sum_{n=0}^{d-1}{ d-1\choose
  n}c^n\beta_n  \,,\qquad\quad
\Lambda_f=\frac{m^d}{m_f^{d-2}}\,c^{2-d}\sum_{n=1}^d{d-1\choose n-1}
c^n\beta_n\,,
\label{Ls}
\ee
where, ${n\choose k}=\frac{n!}{k!(n-k!)}$ is the combinatorial
factor. Note that $\Lambda_g$ does not contain $\beta_d$ while
$\Lambda_f$ is independent of $\beta_0$. 

The consistency of the two equations in (\ref{bg-g_eom}) with each
other then implies,\footnote{The discussion can be easily extended to
  include sources, in which case, (\ref{fcg}) also forces a
  proportionality relation between the two energy-momentum tensors as
  discussed in \cite{HSV3}.}
\be
\Lambda_g=\Lambda_f\,.
\label{LgLf}
\ee
This is a polynomial equation that, generically, determines $c$ in
terms of the parameters of the theory, the exception being the PM case
\cite{Hassan:2012gz} to be discussed later. In particular, the background
equations (\ref{bg-g_eom}) admit de Sitter solutions that are relevant
for the identification of the linear PM theory.

Since (\ref{LgLf}) is homogeneous in the $\beta_k$, the $c$ determined
by it is independent of the overall scale of the $\beta_k$ and will
depend on at most $d+1$ parameters, say $\beta_1/\beta_0,\cdots
\beta_d/\beta_0$ and $\alpha=m_f/m_g$. We are interested in parameter
values that result in real non-zero $c$, as $c$ also appears in other
quantities, like the Fierz-Pauli mass of the fluctuations. In
practice, such parameter ranges can be easily found since, given any
$d$ of the parameters, say, $\{\alpha,
\beta_2/\beta_0,\cdots,\beta_d/\beta_0\}$, one can determine the
allowed range of the remaining one, $\beta_1/\beta_0$, by expressing
it as a function of the real $c$ using (\ref{LgLf}).

\subsection{Linear mass eigenstates}

To determine the mass spectrum of the theory, let us now consider
linear perturbations around the proportional backgrounds,
\be
\gmn=\bgmn+\delta\gmn\,,\quad \fmn=\bfmn +\delta\fmn\,,
\label{fluc}
\ee
with $\bfmn=c^2\bgmn$. It turns out that only around such backgrounds
the fluctuations can be combined into definite mass eigenstates. A
massive mode in a general background is identified through the
appearance of a Fierz-Pauli mass term. On linearizing the equations
of motion (\ref{gf_eom}), one gets (for details, see the appendix)
\begin{align}
&\bar{\mathcal{E}}_{\mu\nu}^{\rho\sigma}\delta g_{\rho\sigma}
-\tfrac{2}{d-2}\Lambda_g\left(\delta\gmn-\frac{1}{2}\bgmn
{\bar g}^{\rho\sigma}\delta g_{\rho\sigma}\right)
+ \tfrac{m^d}{m_g^{d-2}}N\,\bar g_{\mu\lambda}\,\left(\tr(\delta
S)\delta^\lambda_\nu-\delta S^\lambda_{~\nu}\right)=0\,, 
\label{l-g_eom}
\\
&\bar{\mathcal{E}}_{\mu\nu}^{\rho\sigma}\delta f_{\rho\sigma}
-\tfrac{2}{d-2}\Lambda_g\left(\delta\fmn-\frac{1}{2}\bgmn
{\bar g}^{\rho\sigma}\delta f_{\rho\sigma}\right)
-\tfrac{m^d}{m_f^{d-2}}N c^{4-d}\,\bar g_{\mu\lambda}\,
\left(\tr(\delta S)\delta^\lambda_\nu-\delta S^\lambda_{~\nu}\right)=0\,.
\label{l-f_eom}
\end{align}
$\bar{\mathcal{E}}_{\mu\nu}^{\rho\sigma}$ is defined through
(\ref{KO}), now with background metric $\bgmn$, and $N$ is given by
(\ref{N}). $\delta S^\mu_{~\nu}=\frac{1}{2c}\bar g^{\mu\lambda}
(\delta f-c^2 \delta g)_{\lambda\nu}$ enters both equations in the FP
combination, hence we expect to get a massive mode $\delta\Mmn\sim\bar 
g_{\mu\lambda}\delta S^\lambda_{~\nu}$.

The linearized equations are easily diagonalized in terms of a 
massless fluctuation $\delta\Gmn$ and a massive fluctuation
$\delta\Mmn$, 
\begin{align}
&\delta\Gmn=A(c)\left(\delta\gmn+c^{d-4}\alpha^{d-2}\delta\fmn\right)\,,
\label{dG}
\\
&\delta\Mmn=\frac{B(c)}{2c}\left(\delta\fmn-c^2\delta\gmn\right)\,. 
\label{dM}
\end{align}
$A(c)$ and $B(c)$ are normalizations to be determined later and we
have used the notation, 
\be
\alpha=\frac{m_f}{m_g}\,.
\ee
Indeed, adding (\ref{l-g_eom}) and (\ref{l-f_eom}) to cancel the
FP mass term gives a massless spin-2 equation,
\be
\bar{\mathcal{E}}_{\mu\nu}^{\rho\sigma}\delta G_{\rho\sigma}
-\tfrac{2}{d-2}\Lambda_g\left(\delta\Gmn-\frac{1}{2}\bgmn
{\bar g}^{\rho\sigma}\delta G_{\rho\sigma}\right)=0\,.
\label{dG_eom}
\ee
On the other hand, subtracting the right combination of
(\ref{l-g_eom}) and (\ref{l-f_eom}) gives the FP equation
(\ref{linmass}) for the massive spin-2 fluctuation, 
\be
\bar{\mathcal{E}}_{\mu\nu}^{\rho\sigma}\,\delta M_{\rho\sigma}
-\frac{2}{d-2}\Lambda_g\left(\delta \Mmn-\tfrac{1}{2}
\bar{g}_{\mu\nu}\bar{g}^{\rho\sigma}\delta M_{\rho\sigma}\right) 
+\frac{m_{\mathrm{FP}}^2}{2}\,\left(\delta\Mmn-
\bar{g}_{\mu\nu}\bar{g}^{\rho\sigma}\delta M_{\rho\sigma}\right)=0\,. 
\label{dM_eom}
\ee
Here, the FP mass is given by,
\be
m_\mathrm{FP}^2= \frac{m^d}{m_g^{d-2}}
\left(\frac{1+(\alpha c)^{d-2}}{(\alpha c)^{d-2}}\right) 
\sum_{n=1}^{d-1} {{d-2}\choose{n-1}}c^n\beta_n \,.
\label{mfp}
\ee
We emphasize that equations (\ref{dG_eom}) (\ref{dM_eom}) are written
with $\bgmn$ as the background metric. Then the expressions for $\mfp^2$  
(\ref{mfp}) and $\Lambda_g$ (\ref{Ls}) refer to this background metric  
choice.

\subsection{The nonlinear massive and massless fields}

The massless and massive fluctuations can be regarded as perturbations
of some nonlinear fields $\Gmn$ and $\Mmn^G$. If this is the only
criterion, the choice of nonlinear fields is far from unique. However,
we also require that the relation between the nonlinear fields and the
original variables $\gmn$ and $\fmn$ is simple enough that it is
invertible in a useful way \cite{HSV3}. These criteria single out the
following straightforward nonlinear extensions of the mass
eigenstates,
\begin{align}
\Gmn   &= A(c)\left(\gmn + c^{d-4}\alpha^{d-2}\fmn\right)\,,
\label{G}\\
\Mmn^G &=B(c)\left(G_{\mu\rho}S^{\rho}_{~\nu}-c\, G_{\mu\nu}\right)\,.   
\label{MG}
\end{align}
where $S=\sqrt{g^{-1}f}$ and the dimension dependent normalizations
$A$ and $B$ will be fixed later in the context of the PM theory. It is
not claimed that $G$ and $M^G$ propagate respectively $2$ and $5$
degrees of freedom nonlinearly.   

The bimetric action (\ref{Sgf}) can be re-expressed in terms of the
new nonlinear fields as a theory of a massive spin-2 field
$\Mmn^G$ interacting with a massless spin-2 field $\Gmn$. However, it
turns out that $\Gmn$ cannot be coupled to matter in the standard way
without reintroducing the Boulware-Deser ghost and, hence, it cannot
be regarded as the physical gravitational metric \cite{HSV3}. In
other words, the spin-2 mass eigenstates differ from the states
produced by bimetric interactions, as is familiar from other contexts
in particle physics.\footnote{To describe a massive spin-2 field interacting with a
  gravitational metric that is also sourced by conventional matter,
  one should use the fields $\gmn$ and $\Mmn=g_{\mu\lambda}
  S^\lambda_{\,\nu}-c\gmn$ \cite{HSV3}.} However, the theory in
terms of $G$ and $M^G$ is relevant for discussing partial masslessness
which is known to arise only in the absence of matter couplings.

On the proportional backgrounds, the nonlinear fields reduce to, 
\be
\bar G_{\mu\nu} = A(c)\left(1+(\alpha c)^{d-2}\right)\bgmn
\,,\qquad \bar M_{\mu\nu}^G= 0\,.
\label{Ggbg}
\ee
A non-vanishing expectation value for $\Mmn^G$ signals non-proportional
backgrounds and hence parametrizes deviations of the bimetric
solutions from solutions in general relativity.   

The linearized equations (\ref{dG_eom}), (\ref{dM_eom}) are written
with $\bgmn$ as the background metric. To describe the theory entirely
in terms of the massless and massive modes, we rewrite these in terms
of $\bGmn$ using (\ref{Ggbg}),  
\begin{align}
&\tilde{\mathcal{E}}_{\mu\nu}^{\rho\sigma}\,\delta G_{\rho\sigma}
-\frac{2}{d-2}\tilde\Lambda_g\left(\delta G_{\mu\nu}-\tfrac{1}{2}
\bGmn \bar{G}^{\rho\sigma}\delta G_{\rho\sigma} \right)=0\,,
\label{dG-eomG}
\\
&\tilde{\mathcal{E}}_{\mu\nu}^{\rho\sigma}\,\delta M_{\rho\sigma}
-\frac{2}{d-2}\tilde\Lambda_g\left(\delta M_{\mu\nu}-\tfrac{1}{2}
\bGmn\bar{G}^{\rho\sigma}\delta M_{\rho\sigma} \right) +\frac{\tilde
m_{\mathrm{FP}}^2}{2}\,\left(\delta M_{\mu\nu}-\bGmn\bar{G}^{\rho\sigma}
\delta M_{\rho\sigma}\right)=0\,.
\label{dM-eomG}  
\end{align}
Here $\tilde{\mathcal{E}}_{\mu\nu}^{\rho\sigma}$ is given by
(\ref{KO}). In the new background convention, the mass and the
cosmological constant are rescaled to,  
\be
\tilde m_{\mathrm{FP}}^2= \frac{m_{\mathrm{FP}}^2}{A(c)
\left(1+(\alpha c)^{d-2}\right)}\,,
\qquad\qquad
\tilde\Lambda_g= \frac{\Lambda_g}{A(c)
\left(1+(\alpha c)^{d-2}\right)}\,.
\label{tmtL}
\ee

\section{Partial masslessness and Bimetric theory in arbitrary
  dimensions} 

To summarize, on linearizing, the bimetric action leads to the FP
equation for massive spin-2 fields in a cosmological background
(\ref{dM-eomG}), along with equation (\ref{dG-eomG}) for massless
fluctuations of the background itself. Hence it provides a completely
dynamical nonlinear generalization of the FP equation discussed in
section \ref{SeclPM}. In particular, in this setup, one can
investigate partial masslessness beyond the linear level. 

Working in $4$-dimensions, ref.~\cite{Hassan:2012gz} obtained a
criterion to identify the nonlinear PM theory among the set of
ghost-free bimetric theories. Here the considerations are extended to
$d$-dimensions showing that, while linear PM theories exist for any
$d$, nonlinear PM theories based on the bimetric action (\ref{Sgf})
exist only in $d=3$ and $d=4$. This is consistent with the results of
\cite{Joung:2012rv} which shows that, in a 2-derivative theory, spin-2
cubic interactions invariant under the linear PM gauge symmetry exist
only in these dimensions. Furthermore, \cite{Joung:2012rv} shows that
when the $2$-derivative restriction is relaxed, cubic terms with PM
symmetry can be constructed in higher dimensions as well. This case is
considered in the next section.

It should be emphasized that at this stage we do not identify the PM
gauge symmetry in the nonlinear theory, except for a subset of the
transformations around maximally symmetric backgrounds. Hence the claim is
that if a nonlinear theory with a PM gauge symmetry exits, it must
belong to the set of theories identified here. 

\subsection{Linear PM symmetry in bimetric variables}

Since equation (\ref{dM-eomG}) coincides with the FP equation
(\ref{linmass}), it is straightforward to identify the linear PM
theory and the associated gauge symmetry (\ref{gaugesymh}) in bimetric
theory. Hence, when the Higuchi bound is satisfied,  
\be
\tmfp^2=\frac{2}{d-1}\,\tilde\Lambda_g\,,
\label{higuchi-bm}
\ee
equation (\ref{dM-eomG}) develops a new gauge invariance that 
transforms the massive spin-2 fluctuation as $\delta
M_{\mu\nu}\rightarrow \delta M_{\mu\nu}+\Delta(\delta M_{\mu\nu})$,
where  
\be
\Delta(\delta M_{\mu\nu})=
\Big(\nabla_{\mu}\nabla_{\nu}+\frac{2\tilde\Lambda_g}{(d-1)(d-2)}
\bGmn\Big)\xi(x)\,.
\label{gaugesymdM}
\ee
Now we also have an  equation (\ref{dG-eomG}) for the massless
fluctuations $\delta\Gmn$. Since the massless equation has no unusual
symmetries, the above transformation must be supplemented by,
\be
\Delta(\delta G_{\mu\nu})=0
\label{gaugesymdG}
\ee
Using (\ref{dG}) and (\ref{dM}), one obtains the corresponding  
transformations of $\delta g$ and $\delta f$ as,
\begin{align}
\Delta(\delta g_{\mu\nu})= 
-\frac{2}{cB}\,\frac{(\alpha c)^{d-2}}{1+(\alpha c)^{d-2}}\, 
\Delta(\delta \Mmn)\,,
%\nn\\
\qquad
\Delta(\delta f_{\mu\nu})= 
\frac{2c}{B}\,\frac{1}{1+(\alpha c)^{d-2}}\,
\Delta(\delta \Mmn)\,.
\label{gaugesymdgdf}
\end{align}
In any dimension, these equations realize the linear PM gauge
transformations in the bimetric theory around proportional
backgrounds. The associated Higuchi bound (\ref{higuchi-bm}) simply
reduces the number of bimetric parameters by just one and it too can  
be satisfied in any dimension. But we will see that the nonlinear
extension is much more restrictive.

\subsection{Identifying the nonlinear PM theory in $d$ dimensions}  

The gauge transformations (\ref{gaugesymdgdf}) of the linear PM 
theory lead to the criteria for identifying potential
nonlinear PM theories provided one works in a setup that treats the
backgrounds dynamically. The argument is as follows \cite{Hassan:2012gz}: A
nonlinear PM theory must be invariant under an extension of
(\ref{gaugesymdgdf}) that transforms the nonlinear fields as, say,    
\be \gmn \rightarrow g'_{\mu\nu}=\gmn+\Delta\gmn
\,,
\qquad \fmn \rightarrow f'_{\mu\nu}=\fmn+\Delta\fmn\,.  
\ee 
For $g=\bar g+\delta g$, the transformed field also splits as
$g'=\bar g'+\delta g'$, but there is a gauge ambiguity depending on
how $\Delta g$ is split. For instance, a small $\Delta g$ may be
viewed as transforming either the background, $\bar g'=\bar g+\Delta
g$, or the fluctuation, $\delta g'=\delta g+\Delta g$. The same holds 
for $f$.     

The implication of this for the linearized bimetric theory is that the
variation $\Delta(\delta g)$ of $\delta g$ (\ref{gaugesymdgdf}) can,
alternatively, be regarded as a 
transformation of the background $\bar g$ to $\bar g'=\bar g+
\Delta(\delta g)$, keeping $\delta g$ unchanged. The same holds for
$\fmn$. Let's adopt this latter point of view. Now, for a generic
gauge parameter $\xi(x)$ in (\ref{gaugesymdM}), the new ($\bar g',
\bar f'$) are not proportional backgrounds and hence are not dS 
metrics, in which case not much is known about partial masslessness.
To keep to dS backgrounds we restrict $\Delta(\delta g)$ and
$\Delta(\delta f)$ to constant $\xi(x)=\xi_0$ and rename them to
$\Delta_0(\bar g)$, $\Delta_0(\bar f)$ to emphasize that they
transform $(\bar g,\bar f)$ rather than $(\delta g,\delta f)$. Then,
given $\bar f=c^2\,\bar g$, the new backgrounds are related through,  
\be
\bar f'=c'^2(\xi_0)\, \bar g'\,, 
\label{f'c'g'}
\ee 
and the dS preserving gauge transformations at background level are,  
\be
\bar g'=\bar g+\Delta_0(\bar g)\,,\qquad 
\bar f'=\bar f+\Delta_0(\bar f)\,,\qquad
c'=c+ \Delta_0(c)\,.  
\label{gaugesymBG}
\ee
The $\Delta_0$-variations above are obtained from
(\ref{gaugesymdgdf}), (\ref{gaugesymdM}) and (\ref{f'c'g'}) for
constant $\xi(x)=\xi_0$, 
\be
\Delta_0(\bg)=-\frac{2}{c}(\alpha c)^{d-2}\,\bg\,\lambda_d
\,\xi_0\,,\quad 
\Delta_0(\bar f)=2c\,\bg\,\lambda_d \,\xi_0\,,\quad
\Delta_0(c)=\left[1+(\alpha c)^{d-2}\right]\lambda_d\,\xi_0\,,
\label{Delta0s}
\ee
where, $\lambda_d$ stands for,
\be
\lambda_d=\frac{A}{B} \frac{2\tilde\Lambda_g}{(d-1)(d-2)}\,.
\label{ld}
\ee
$A$ and $B$ are the normalizations of the nonlinear massless and
massive fields $\Gmn$ and $\Mmn^G$ to be determined later. 
 
Thus, the conclusion is that if a nonlinear theory with PM gauge
symmetry exits, its equations of motion on proportional backgrounds
must be invariant under the transformations (\ref{gaugesymBG}). This
requirement puts constraints on the bimetric theory (\ref{Sgf}) and
singles out the potential nonlinear PM cases. The criteria for the
invariance are listed below.   
\begin{enumerate}
\item Recall that for proportional backgrounds $\bar f=c^2\bg$, the
  bimetric equations of motion imply $\Lambda_g=\Lambda_f$
  (\ref{LgLf}), explicitly, 
\be 
\sum_{n=0}^{d-1}{ d-1\choose
    n}c^n\beta_n =\,(\alpha c)^{2-d}\sum_{n=1}^d{d-1\choose n-1}
  c^n\beta_n\,.
\label{Lg=Lf}
\ee 
Generically, this fixes $c=c(\alpha,\beta_i)$ and excludes the
possibility of invariance under transformations (\ref{gaugesymBG}) that
require changing $c$. Thus the nonlinear theory can be invariant under
(\ref{gaugesymBG}) only for special values of ($\beta_n,\alpha$) for
which (\ref{Lg=Lf}) leaves $c$ undetermined. This is the necessary
condition for the existence of nonlinear PM theories for these
($\beta_n,\alpha$) values. 
\item When the theory is expressed in term of the massive and massless  
  fields $\Mmn^G$ and $\Gmn$, the FP mass and the cosmological
  constant take the form (\ref{tmtL}), or explicitly,
\begin{align}
\tilde m_{\mathrm{FP}}^2 &=\frac{m^d}{m_g^{d-2}}\,
\frac{(\alpha c)^{2-d}}{A(c)}\,
\sum_{n=1}^{d-1} {{d-2}\choose{n-1}}c^n\beta_n \,,
\label{tm}\\[.2cm]
\tilde\Lambda_g&= \frac{m^d}{m_g^{d-2}}
\frac{\left(1+(\alpha c)^{d-2}\right)^{-1}}{A(c)} 
\sum_{n=0}^{d-1}{ d-1\choose
  n}c^n\beta_n  \,.
\label{tL}
\end{align}
Since the gauge transformations in the nonlinear PM theory involve
shifts of $c$, then $\tmfp^2$ and $\tilde\Lambda_g$, as well as the
effective couplings, must become $c$-independent. 
\item The gauge symmetry of the linear PM theory leaves the massless 
  fluctuation $\delta\Gmn$ invariant. Hence at the background level
  too, transformations (\ref{gaugesymBG}) must keep the corresponding
  nonlinear background $\bGmn$ invariant. This requirement determines
  the normalization $A(c)$ of $\Gmn$. Indeed from (\ref{Ggbg}) and
  (\ref{gaugesymBG}) one can compute, 
\be
\Delta_0 \bGmn=\bgmn A\,\left[1+(\alpha c)^{d-2}\right]\,\Delta_0(c)
\,\frac{d}{dc}\ln\left(A\left[1+(\alpha c)^{d-2}
\right]^{\frac{d-4}{d-2}}\right)\,=0\,,
\ee
which fixes the normalization, upto a $c$-independent factor, as,
\be
A(c)=\left[1+(\alpha c)^{d-2}\right]^{\frac{4-d}{d-2}}\,,
\label{A}
\ee
\end{enumerate}
Then, the background massless and massive fields become,
\be
\bar G_{\mu\nu} = 
\left[1+(\alpha c)^{d-2}\right]^{\frac{2}{d-2}}\,
\bgmn
\,,\qquad \bar M_{\mu\nu}^G= 0\,.
\label{Ggbg'}
\ee
As discussed below, the first criterion is satisfiable only in $3$ and 
$4$ dimensions (not counting $d<3$ which has trivial dynamics). Then
in these cases, the second criterion is satisfied only for the
normalizations $A(c)$ determined by the third criterion. Hence, the
potential nonlinear PM theories in $3$ and $4$ dimensions are
manifestly invariant under the dS preserving part of the PM gauge
transformations. Let us now consider these cases separately. 

\subsection{Nonlinear PM theory for d=3, 4}

In $3$ dimensions the condition (\ref{Lg=Lf}) that generically
determines the $c$ in $f=c^2g$ reads  
\be
(\alpha\beta_0-\beta_1)+2(\alpha\beta_1-\beta_2)c+(\alpha\beta_2-
\beta_3)c^2=0\,. 
\ee
It leaves $c$ undetermined only when the coefficients of all powers
of $c$ vanish,  
\be
\beta_1=\alpha\beta_0\,,\qquad
\beta_2=\alpha\beta_1\,,\qquad
\beta_3=\alpha\beta_2\,,
\label{PMbs3}
\ee
or simply, $\beta_n=\alpha^n\beta_0$ for $n=1,2,3$. Only for these 
parameter values the restricted gauge transformations 
(\ref{gaugesymBG}) lead to new proportional backgrounds $f'=c'^2g'$ 
with $c'\neq c$, without violating the bimetric equations of motion.  
For the above parameter values, the mass and cosmological constant
(\ref{tm}, (\ref{tL}) measured in the metric $\bGmn$ become,
\be
\tmfp^2=\tilde\Lambda_g = \frac{m^3}{m_f}\beta_1\,.
\ee
Thus they satisfy the Higuchi bound and are indeed independent of $c$
and hence gauge invariant, but only for the normalization
$A=\left[1+(\alpha c)\right]$ given by (\ref{A}).  

In $4$ dimensions (first considered in \cite{Hassan:2012gz}), the condition  
(\ref{Lg=Lf}) becomes,
\be
\beta_1+\left(3\beta_2-\alpha^2\beta_0\right)c+\left(3\beta_3-
3\alpha^2\beta_1\right)c^2+\left(\beta_4-3\alpha^2\beta_2\right)c^3
+\alpha^2\beta_3c^4 =0\,.
\ee
This leaves $c$ undetermined for 
\be
\beta_1=\beta_3=0\,,\qquad
\alpha^4\beta_0=3\alpha^2\beta_2=\beta_4\,. 
\label{PMbs4}
\ee
Then the nonlinear PM theory must correspond to this choice of
parameters. For these parameters, the mass and cosmological
constant satisfy the Higuchi bound,
\be
\tmfp^2=\frac{2}{3}\tilde\Lambda_g =2\,\frac{m^4}{m_f^2}\,\beta_2\,.
\ee
Again, they are independent of $c$, and hence gauge invariant,
precisely for the normalization $A=1$ (\ref{A}) that also renders
$\bGmn$ invariant.  

The bimetric action (\ref{Sgf}) with the $\beta$'s given above 
satisfies the necessary conditions for being a nonlinear PM
theory for $d=3,4$. The PM gauge symmetry of the nonlinear action
(\ref{Sgf}) is not yet known. But the discussion shows that, at least
for proportional backgrounds, the nonlinear equations are invariant
under the dS preserving subset of the PM gauge transformations 
(\ref{gaugesymBG}). In fact, this subset can be further extended to
the nonlinear dS preserving transformations, 
\be 
c'=c+a\,,\qquad \bgmn'=\left[\frac{1+(\alpha c)^{d-2}}
{1+(\alpha (c+a))^{d-2}}\right]^{\frac{2}{d-2}}\, 
\bgmn \,,
\ee
where $a$ is a finite parameter and the transformation of $\bar f$
follows from $\bar f=c^2 \bar g$. To linear order in $a=\Delta_0(c)$,
this reduces to (\ref{gaugesymBG}). The above transformations keep the 
background $\bar G_{\mu\nu}$ (\ref{Ggbg}) unchanged and also leave the 
background $\gmn$ and $\fmn$ equations of motion (\ref{bg-g_eom}) 
invariant.  

\subsection{The quadratic action}

As a last check, let us consider the quadratic action for the
fluctuations $\delta\Gmn$ and $\delta\Mmn$ given by
(\ref{dG}),(\ref{dM}) and (\ref{A}) with $B(c)=1$, that is,
\begin{align}
&\delta\Gmn=\left[1+(\alpha c)^{d-2}\right]^{\frac{4-d}{d-2}}
\left(\delta\gmn+c^{d-4}\alpha^{d-2}\delta\fmn\right)\,,
\label{dG2} \\
&\delta\Mmn=\frac{1}{2c}\left(\delta\fmn-c^2\delta\gmn\right)\,. 
\label{dM2}
\end{align}
This action is needed to determine the effective Planck
masses of $\delta\Gmn$ and $\delta\Mmn$ and to verify that they are
indeed independent of $c$ and hence gauge invariant. Expanding the
bimetric action (\ref{Sgf}) to second order and expressing the result
in terms of the above combinations one obtains,
\begin{align}
S^{(2)}&= m_g^{d-2}\int\td^dx 
\sqrt{\bar G}\biggl[
-\delta G\tilde{\mathcal{E}}\delta G+\frac{\tilde\Lambda_g}{(d-2)}
\bigg(\mathrm{Tr}[\delta G^2]-\frac1{2}\mathrm{Tr}[\delta G]^2\bigg)
\nn\\
&-\delta M\tilde{\mathcal{E}}\delta M-\frac{\tilde\Lambda_g}{(d-2)}
\bigg(\mathrm{Tr}[\delta M^2]-\frac1{2}\mathrm{Tr}[\delta M]^2\bigg)
+\frac{\tilde m_{\mathrm{FP}}^2}{4} \bigg(\mathrm{Tr}[\delta M]^2-
\mathrm{Tr}[\delta M^2]\bigg) 
\biggr].
\label{quadS}
\end{align}
Here, $\tilde m_{\mathrm{FP}}^2$ and $\tilde\Lambda_g$ are given by
(\ref{tm}), (\ref{tL}) which become independent of $c$ for the PM
parameter values. Also note that the Planck masses are independent of
$c$ for the correct normalization in (\ref{dG2}) determined by
other considerations. Furthermore, the quadratic action fixes $B(c)=1$
since otherwise, the $\delta M$ sector will have a $c$-dependent
coupling. 

In the appendix we outline the structure of the complete nonlinear
action expressed in terms of $\Gmn$ and $M^G_{\mu\nu}$. This is useful
for computing the cubic and higher order interactions of the PM field
to be compared with the explicit cubic vertex computations in
\cite{Zinoviev:2006im, Joung:2012rv, Deser:2012qg}. We do not attempt
such a comparison here. 

\subsection{Absence of nonlinear PM theory in d$>$4}

The necessary condition for the existence of nonlinear PM theories is
that (\ref{Lg=Lf}) leaves $c$ undetermined. On relabeling
the sum on the right hand side, this equation can be recast as, 
\be
\left(\alpha^{d-2}\sum_{n=0}^{d-1}A_n\,\beta_n
-\sum_{n=3-d}^{2} B_{n+d-2}\,\beta_{n+d-2}\right)\,c^n=0\,,
\ee
where, 
\be
A_n= {d-1\choose n}\,,\qquad B_n= {d-1\choose n-1}\,.
\ee
To leave $c$ undetermined, the coefficients of each power of $c$ must 
vanish separately. For $d>4$, this gives the following 3 sets of
equations:
\begin{align}
3-d \leq  n \leq -1 & : \quad \beta_{n+d-2}=0\,,\\  
0\leq n \leq 2  & : \quad 
\alpha^{d-2}A_n\,\beta_n-B_{n+d-2}\,\beta_{n+d-2}=0\,, \\
3\leq < n \leq d-1  & : \quad \beta_n=0\,.
 \end{align}
The first set implies that $\beta_m=0$ for $1\leq m\leq d-3$ whereas
the last set implies that $\beta_m=0$ for $3\leq m\leq d-1$. Then the
second set implies $\beta_0=\beta_d=0$. Hence, $\beta_n=0$ for
$n=0,\cdots, d$.

To recapitulate, the bimetric action (\ref{Sgf}), when linearized
around proportional backgrounds, leads to the massive FP equation
(\ref{dM-eomG}). On the Higuchi bound (\ref{higuchi-bm}), this gives a
linear PM theory in any dimension. The consistency of the
dS-preserving part of the linear PM gauge symmetry with dynamical
backgrounds provided the necessary (bot not sufficient) condition for
the existence of nonlinear PM theories. This criterion was satisfied
in $3$ and $4$ dimensions but not for $d>4$.

\section{Partial Masslessness with Lanczos-Lovelock terms}

Our nonlinear results are consistent with the observation in
\cite{Joung:2012rv} that in a $2$-derivative theory, cubic spin-2
interactions with PM gauge symmetry exist only in $3$ and $4$
dimensions. However, \cite{Joung:2012rv} also finds that in the
presence of higher derivative terms, cubic PM interactions exist even
for $d>4$. These higher derivative terms reduce to 2-derivative terms
in $d=4$. This hints that the nonlinear higher derivative interactions
correspond to Lanczos-Lovelock (LL) terms which will also preserve the
unitarity of the bimetric theory. This section considers the
LL extension of the bimetric theory from the point of view of the PM
analysis. Such a theory has been considered in \cite{Paulos:2012xe} in
a different context. Recently, in \cite{Sisman:2012rc} the spectrum of
Einstein-Hilbert gravity with LL terms was studied in dS backgrounds.
Here, this analysis is generalized to the bimetric theory and used to
investigate the existence of PM theories beyond $d=4$.

To summarize, it is shown that, for $\bar f=c^2\bar g$ and around dS
backgrounds, the structure of massless and massive fluctuations in
bimetric theory with LL terms is exactly the same as in pure bimetric
theory, though with modified parameters. Hence the PM analysis
previously developed for the pure bimetric case also applies here. In
particular, the linearized theory has PM gauge symmetry on the Higuchi
bound. Then for a nonlinear extension of the theory to exist, the
necessary condition is that the equations of motion must leave $c$
undetermined. This condition is easily implemented if one obtains a
polynomial equation for $c$.

For $d\leq 4$ the LL terms are of no consequence and the pure
bimetric analysis goes through. The first new contribution is from the
quadratic LL term for $d\geq 5$ and this remains the only contribution
for $d=5,6$. In this case, one can easily obtain a polynomial equation
for $c$ and determine the potential PM theory.  The cubic LL term
starts contributing for $d>6$ but in this case one does not
automatically obtain a polynomial equation for $c$. However, using a
decomposition into the Gr\"obner basis one again obtains a polynomial
equation for $c$, as discussed in subsection \ref{cubicll}.  The
analytical derivation of the PM theory in $d=5$ is presented in
subsection \ref{d5analytical}, whereas the PM theories for $d=6,7,8$
are determined with the help of a computer and are discussed in
subsection \ref{dgrt5}. Beyond this, the analysis gets more involved
and is not attempted here.

\subsection{Bimetric theory with Lanczos-Lovelock terms in dS
  spacetimes}

Adding higher order LL invariants $\mathcal{L}_{(n)}$ for both $\gmn$
and $\fmn$ to the bimetric action gives,
\begin{align}\label{actll}
S =& m_g^{d-2}\int\td^dx\sqrt{g}\left[R(g)+\sum_{n=2}^{[d/2]}l^g_n
\mathcal{L}_{(n)}(g)\right]+m_f^{d-2}\int\td^dx\sqrt{f}\left[R(f)+
\sum_{n=2}^{[d/2]}l^f_n\mathcal{L}_{(n)}(f)\right]\nn\\
&-2m^d\int\td^dx\sqrt{g}~V\left(\sqrt{g^{-1}f};\beta_n\right)\,,
\end{align}
where $l^g_n$ and $l^f_n$ are coupling constants of mass dimension
$-2(n-1)$. 
The sum terminates at $[d/2]$ (integer part of) since the Lovelock
terms vanish identically when $d<2n$ and are topological invariants
when $d=2n$. Our conventions for the Lovelock invariants are,
\be
\mathcal{L}_{(n)}=\frac1{2^n}\delta^{\mu_1\nu_1\dots\mu_n\nu_n}_{\alpha_1
 \beta_1\dots\alpha_n\beta_n}\prod_{r=1}^n
R^{\alpha_r\beta_r}_{\ph{\alpha_r\beta_r}\mu_r\nu_r}\,,\quad
\delta^{\mu_1\nu_1\dots\mu_n\nu_n}_{\alpha_1\beta_1\dots\alpha_n\beta_n}
=\frac1{n!}\delta^{\mu_1}_{[\alpha_1}\delta^{\nu_1}_{\beta_1}
\cdots\delta^{\mu_n}_{\alpha_n}\delta^{\nu_1}_{\beta_1]}\,.
\ee 
Note that the action (\ref{actll}) is invariant under simultaneous
interchange of,  
\be
\label{intsym}
\alpha^{\frac{2-d}{2}}\gmn\longleftrightarrow\alpha^{\frac{d-2}{2}}\fmn\,, 
\qquad\beta_n\longrightarrow\alpha^{2n-d}\beta_{d-n}\,,
\qquad l^f_n \longleftrightarrow \alpha^{2-2n}l^g_n\,.
\ee
The equations of motion obtained from this action read,
\begin{align}
&R_{\mu\nu}(g)-\tfrac1{2}\gmn R(g)+\sum_{n=2}^{[d/2]}l_n^g
\mathcal{G}^{(n)}_{\mu\nu}(g)+\tfrac{m^d}{m_g^{d-2}}V^g_{\mu\nu}=0\,,
\label{eom-gLL}\\ 
&R_{\mu\nu}(f)-\tfrac1{2}\gmn R(f)+\sum_{n=2}^{[d/2]}
l_n^f\mathcal{G}^{(n)}_{\mu\nu}(f)+\tfrac{m^d}{m_f^{d-2}}V^f_{\mu\nu} 
= 0\,,
\label{eom-fLL}
\end{align}
where the Lovelock tensors $\mathcal{G}^{(n)}_{\mu\nu}$ appear as the
result of varying the Lovelock invariants.  

We are interested in the maximally symmetric solutions of
the above equations that allow for massive spin-2 excitations. In the
bimetric setup, these belong to the class of proportional
backgrounds $\fmn=c^2\gmn$. For a maximally symmetric spacetime with
cosmological constant $\lambda$, the curvatures are,
\be
\label{maxsymr}
R_{\mu\nu\rho\sigma}(g) = \frac{2\lambda}{(d-1)(d-2)}
\left(g_{\mu\rho}g_{\nu\sigma}-g_{\nu\rho}g_{\mu\sigma}\right)\,,\quad
R_{\mu\nu}(g) = \frac{2\lambda}{d-2}\gmn\,,\quad R(g) = 
\frac{2d\lambda}{d-2}\,,
\ee
with the corresponding equations for $\fmn$. With this ansatz it is
enough to consider the traces of the equations of motion
(\ref{eom-gLL}) and (\ref{eom-fLL}) to obtain two equations that
generically determine $\lambda$ and $c$ in terms of the parameters of
the theory. The computation is simplified by,
\be
\label{LovG_LovL}
g^{\mu\nu}\mathcal{G}^{(n)}_{\mu\nu}(g)=
\frac{2n-d}{2}\mathcal{L}_{(n)}(g)\,,
\ee
which allows us to work with the Lovelock invariants, rather than
the corresponding Lovelock tensors. Furthermore, for the proportional
ansatz we have, 
\be\label{Lovf_Lovg}
	\mathcal{L}_{(n)}(f) = c^{-2n}\mathcal{L}_{(n)}(g)\,.
\ee
Now, for the ansatz (\ref{maxsymr}), the Lovelock invariants become, 
\be
\label{llbg}
\mathcal{L}_{(n)} = N_n(d)\lambda^n\,,\qquad \text{with}\quad N_n(d) 
= \frac{2^n d!}{(d-1)^n(d-2)^n(d-2n)!}\,.
\ee
Using the above relations, the traced equations of motion
(\ref{eom-gLL}) and (\ref{eom-fLL}) give,
\begin{align}
\lambda +\sum_{n=2}^{[d/2]}l^g_n\,\frac{d-2n}{2d}N_n(d)\lambda^n
-\Lambda_g&=0\,,
\label{trbgeqg}\\
\lambda +\sum_{n=2}^{[d/2]}c^{2-2n}l^f_n\,\frac{d-2n}{2d}N_n(d)
\lambda^n-\Lambda_f&=0\,.
\label{trbgeqf}
\end{align}
The contributions $\Lambda_{g,f}$ from the bimetric potential are
still given by (\ref{Ls}). In general, these equations determine the
cosmological constant $\lambda$ and the proportionality constant $c$.
Hence, they specify the maximally symmetric solutions in bimetric theory 
with LL terms.

Now, we consider the spectrum of fluctuations in the above dS
backgrounds. The analysis is greatly simplified using results of
\cite{Sisman:2012rc}, which considered Einstein-Hilbert gravity with LL terms,
$S_{EH+LL}$. For a single metric $\gmn=\bgmn+\delta\gmn$ described by
$S_{EH+LL}$, \cite{Sisman:2012rc} showed that the quadratic action for
$\delta\gmn$ in an (A)dS background $\bgmn$ was exactly the same as 
the quadratic action in pure Einstein-Hilbert gravity in the same
background but with a modified, effective, Planck mass. Applied to the
bimetric action with LL terms (\ref{actll}) this implies that at the 
quadratic level, the theory has the same structure as the pure
bimetric theory (\ref{Sgf}) discussed earlier, but now with modified
Planck masses,  
\begin{align}
\bar{m}_g^{d-2} &= m_g^{d-2}\Big[1+(d-3)!\sum_{n=2}^{[d/2]}
\frac{n(d-2n)}{(d-2n)!}\left(\frac{2\lambda}{(d-1)(d-2)}\right)^{n-1}
l_n^g\Big]\,,
\label{bmg}\\	
\bar{m}_f^{d-2} &= m_f^{d-2}\Big[1+(d-3)!\sum_{n=2}^{[d/2]}
\frac{n(d-2n)}{(d-2n)!}\left(\frac{2\lambda}{(d-1)(d-2)}\right)^{n-1}
c^{2-2n}l_n^f\Big]\,. \label{bmf}
\end{align}
This implies that, just as before, the spectrum of fluctuations
consists of massive and massless spin-2 modes in a dS background with
cosmological constant $\lambda$. In particular, the Fierz-Pauli mass
has the same form as in (\ref{mfp}), but now involves the modified 
Planck masses,        
\be
\label{llfp}
m_\mathrm{FP}^2=\frac{m^d}{\bar{m}^{d-2}_g}\left(1+\left(c
\frac{\bar{m}_f}{\bar{m}_g}\right)^{2-d}\right)\sum_{k=1}^{d-1}{d-2
\choose k-1}c^k\beta_k\,.  
\ee

Since the massive mode satisfies the FP equation, it is obvious that
the linear theory will exhibit partial masslessness at the Higuchi
bound,  $m^2_\mathrm{FP}=\frac{2}{d-1}\lambda$. Below we show that in
the presence of the LL terms, nonlinear potentially PM theories can be
found even for $d>4$. Before this, let us look at the quadratic and
quartic LL terms in more detail.

\subsection{The quadratic Lanczos-Lovelock term}\label{quadraticll}

We now apply the general results of the previous section to the
quadratic LL term which is the Gauss-Bonnet combination, 
\be
\mathcal{L}_{(2)} = R^2 -4R_{\mu\nu}R^{\mu\nu}+R_{\mu\nu\rho\sigma}
R^{\mu\nu\rho\sigma}\,.
\ee
This contribution is trivial for $d\leq 4$ but modifies the bimetric
theory for $d>4$. For a maximally symmetric spacetime with
cosmological constant $\lambda$ it reduces to 
\be
\mathcal{L}_{(2)} = \frac{4d(d-3)}{(d-1)(d-2)}\lambda^2\,.
\ee
Then the traced bimetric equations of motion on maximally symmetric, 
proportional backgrounds become 
\begin{align}
\lambda+\frac{2(d-3)(d-4)}{(d-1)(d-2)}l_2^g\lambda^2-\Lambda_g 
&= 0\,,\label{gtreq1}\\
\lambda+\frac{2(d-3)(d-4)}{(d-1)(d-2)}c^{-2}l_2^f\lambda^2-\Lambda_f
&= 0\,,\label{ftreq1}
\end{align}

The above equations determine $\lambda$ and $c$ in terms of the
parameters of theory. For the purpose of PM analysis, we prefer that
$c$ is determined by a polynomial equation. For this, the solution
for $\lambda$ can be easily obtained in the appropriate form on
multiplying the $f$-equation by $c^{2}l_2^g/l_2^f$ and subtracting it
from the $g$-equation. This gives,
\be
\label{smalll}
\lambda=\frac{c^2l_2^g\Lambda_f-l_2^f\Lambda_g}{c^2l_2^g-l_2^f}\,.
\ee
Then plugging this expression into the difference of (\ref{gtreq1})
and (\ref{ftreq1}) gives 
\be
\label{eqcind2}
\frac{2(d-3)(d-4)}{(d-1)(d-2)}l_2^g(l\Lambda_g-c^2\Lambda_f)^2+
(c^4-lc^2)(\Lambda_f-\Lambda_g)=0\,,
\ee
where $l\equiv l_2^f/l_2^g$. This is the desired polynomial equation
for $c$ which is a generalization of the previously considered
condition $\Lambda_f=\Lambda_g$. In order to arrive at the PM
candidate, we will demand that this equation leaves $c$ undetermined.  
 
The effective Planck masses in terms of which the FP mass should be
expressed now read,
\be
\bar{m}_g^{d-2}=m_g^{d-2}\left(1+\frac{4(d-3)(d-4)}{(d-1)(d-2)}l_2^g
\lambda\right)\,,\qquad	 
\bar{m}_f^{d-2} = m_f^{d-2}\left(1 + \frac{4(d-3)(d-4)}{(d-1)(d-2)}
c^{-2}l_2^f\lambda\right)\,.
\ee
These results are sufficient to investigate the most general PM
theories in dimension $d=5$ and $d=6$ where only the quadratic
LL term can be added to the bimetric action. 

\subsection{The cubic Lanczos-Lovelock term}\label{cubicll}

The cubic LL invariant is given by
\begin{align}
\mathcal{L}_{(3)} =& R^3 -8R^{\mu\nu\rho\sigma}
R_{\mu\ph\tau\rho}^{\ph\mu\tau\ph\rho\gamma}R_{\nu\tau\sigma\gamma}
+4R^{\mu\nu\rho\sigma}R_{\mu\nu}^{\ph\mu\ph\nu\tau\gamma}
R_{\rho\sigma\tau\gamma} -24
R^{\mu\nu}R^{\rho\sigma\tau}_{\ph{\rho\sigma\tau}\mu}R_{\rho\sigma\tau\nu}
\nn\\
&+3RR_{\mu\nu\rho\sigma}R^{\mu\nu\rho\sigma}
+24R^{\mu\nu}R^{\rho\sigma}R_{\mu\rho\nu\sigma}
+16R^{\mu\nu}R^\rho_\mu R_{\nu\rho}-12RR_{\mu\nu}R^{\mu\nu}\,,
\end{align}
which on maximally symmetric backgrounds reduces to
\be
\mathcal{L}_{(3)} = \frac{8d(d-3)(d-4)(d-5)}{(d-1)^2(d-2)^2}\lambda^3\,.
\ee
The trace of the equations of motion on proportional, maximally
symmetric backgrounds then gives, 
\begin{align}
\lambda+\frac{2(d-3)(d-4)}{(d-1)(d-2)}l_2^g\lambda^2+
\frac{4(d-3)(d-4)(d-5)(d-6)}{(d-1)^2(d-2)^2}l_3^g\lambda^3-
\Lambda_g &= 0\,,
\label{gtreq}\\
\lambda+\frac{2(d-3)(d-4)}{(d-1)(d-2)}c^{-2}l_2^f\lambda^2+
\frac{4(d-3)(d-4)(d-5)(d-6)}{(d-1)^2(d-2)^2}c^{-4}l_3^f\lambda^3-
\Lambda_f &= 0\,.
\label{ftreq}
\end{align}
Subtracting off the cubic term will result in a quadratic equation for
$\lambda$ and hence the solution will involve a square root. However,
in order to apply our method for determining the PM candidate, we need
to arrive at a polynomial equation in $c$. It turns out that this can
still be achieved by decomposing the equations into a Gr\"obner basis
of equivalent polynomials instead. In the cases we have been able to
reduce in this way it turns out that we then get a linear equation for
$\lambda$ together with a consistency condition on the coefficients of
the original polynomials which can be converted into a polynomial
equation in $c$. 
Here, for simplicity, we present the results for the Gr\"obner
reduction after having imposed the symmetry relation
$l^f_n=\alpha^{2-2n}l^g_n$ which also covers the case of PM theories
(although later the PM analysis is performed starting with the most
general set of parameters). We then obtain a linear equation for
$\lambda$ with the solution  
\begin{align}
\lambda =\frac{n_3(1-q^2)(1+q^2)^2\Delta_q\Lambda-n_2^2q^2(1-q^2)
[\Delta_q\Lambda +2q^2(\Lambda_g-\Lambda_f)]-
n_2n_3(\Delta_q\Lambda)^2}{(1-q^2)
[n_3(1-q^2)(1+q^2)^3-n_2^2q^2(1-q^4-n_2(\Lambda_g-q^2\Lambda_f))]}\,,
\end{align}
where, we have used the following notion, 
\be
\Delta_q\Lambda=\Lambda_g-q^4\Lambda_f\,,\quad q=\alpha c\,,
\quad n_k = \frac{d-2n}{2d}N_k(d)l^g_k\,.
\ee
A consistency condition on the coefficients of the original
polynomials is,  
\begin{align}
&n_3^2(\Lambda_g-q^4\Lambda_f)^3+n_2^2q^4(1-q^2)\bigl[q^4(\Lambda_g
-\Lambda_f-n_2\Lambda_f^2)-n_2\Lambda_g^2-q^2(\Lambda_g-\Lambda_f-2n_2 
\Lambda_g\Lambda_f)\bigr]
\nn\\
&-n_3q^2(1-q^2)\bigl[q^6(1+3n_2\Lambda_f)(\Lambda_g-\Lambda_f)+
q^8(\Lambda_g-\Lambda_f-n_2\Lambda_f^2)-n_2\Lambda_g^2\nn\\
&\qquad\qquad\qquad\quad-q^2(1+3n_2\Lambda_g)(\Lambda_g-\Lambda_f)
-q^4(\Lambda_g-\Lambda_f-2n_2\Lambda_g\Lambda_f)\bigr]=0\,.
\end{align}
On multiplying by a factor $c^{3(d-7)}$, one obtains a polynomial
equation in $c$. The requirement that this leaves $c$ undetermined
constrains the $\beta_n$ and the $l^f_n$ as given in Table \ref{table}.
With these results for the cubic term we can investigate PM theories
up to $d=8$ in full generality.  

\subsection{Partial Masslessness with LL terms in $d=5$}\label{d5analytical}

Let us now consider the nonlinear PM candidate with higher derivative
interactions in five dimensions where only the quadratic LL term is
non-vanishing. This theory is parametrized by the six $\beta_n$, along
with $l_2^g, l_2^f$ and the two Planck masses $m_g, m_f$. In this
case, (\ref{Ls}) gives,
\begin{align}
\Lambda_g&=\frac{m^5}{m_g^3}(\beta_0+4c\beta_1+6c^2\beta_2
+4c^3\beta_3+c^4\beta_4)\,,\\
\Lambda_f&=\frac{m^5}{m_g^3}(\alpha c)^{-3}(c\beta_1
+4c^2\beta_2+6c^3\beta_3+4c^4\beta_4+c^5\beta_5)\,. 
\end{align}
Inserting these into (\ref{eqcind2}) gives a polynomial with nine
different powers in $c$. Demanding it to be independent of $c$ by
setting the coefficient of each power of $c$ to zero therefore gives
nine conditions on the parameters that can be explicitly solved. 

To simplify the presentation, here we first impose the extra
requirement that the parameters of the PM theory must correspond to a
fixed point of the interchange symmetry (\ref{intsym}), or else one
ends up with two sets of nonlinear PM theories related by
(\ref{intsym}). Of course, we have checked that the unique nontrivial
solution is also obtained without making this ansatz. Then this
requirement gives,  
\be
\label{symans}
\beta_n=\alpha^{2n-d}\beta_{d-n}\,,\qquad l^f_n = \alpha^{2-2n}l^g_n\,,
\ee
or, explicitly, $\beta_5=\alpha^5\beta_0, \beta_4=\alpha^3\beta_1,
\beta_3=\alpha\beta_2, l_2^f=\alpha^{-2}l_2^g$.
Imposing these, it follows that the nine equations reduce to three
independent ones,   
\begin{align}
(\alpha\beta_0-\beta_1)^2-3\kappa\alpha\beta_1&=0\,,\\
2(\alpha\beta_0-\beta_1)(\alpha\beta_1-\beta_2)
-3\kappa\alpha\beta_2&=0\,,\\
16(\alpha\beta_1-\beta_2)^2-18\kappa\alpha^2\beta_2
+3\kappa\alpha^4\beta_0+3\kappa\alpha^3\beta_1&=0\,,
\end{align}
where $\kappa\equiv\frac{m_g^3}{l_2^gm^5}$ is a dimensionless
parameter. The structure suggests the following definitions,
\be
\beta_0\equiv-3\kappa b_0\,,\qquad \beta_1\equiv
-3\kappa\alpha b_1\,,\qquad \beta_2\equiv-3\kappa\alpha^2 b_2\,.
\ee
Then the system reduces to the three equations,
\be
(b_0-b_1)^2+b_1=0,\quad  2(b_0-b_1)(b_1-b_2)+b_2=0,
\quad 16(b_1-b_2)^2-b_0-b_1+6b_2=0,
\ee
which are solved by $b_0=\frac{5}{36}, b_1= -\frac{1}{36},
b_2=\frac{1}{72}$, or, 
\begin{align}
\beta_0&=-\frac{5}{12}\kappa,\quad\beta_1=\frac{\alpha}{12}\kappa,
\quad\beta_2=-\frac{\alpha^2}{24}\kappa,\quad
\beta_3=-\frac{\alpha^3}{24}\kappa,\nn\\
\beta_4&=\frac{\alpha^4}{12}\kappa,\quad\beta_5
=-\frac{5\alpha^5}{12}\kappa,\quad l_2^f=\alpha^{-2}l_2^g\,. 
\end{align}
These determine all the $\beta_n$ and $l_2^f$ in terms of $l_2^g$,
$m_g$ and $m_f$. 
Inserting these into the solution for $\lambda$ in equation (\ref{smalll}),  
and into the FP mass (\ref{llfp}) gives the linear Higuchi bound,
\be
m_\mathrm{FP}^2=\frac{1}{4l_2^g}(-1+\alpha c-\alpha^2c^2)
=\frac{1}{2}\lambda\,.
\ee

\subsection{Partial masslessness with LL terms in $d>5$}\label{dgrt5}

In $d=6$, the cubic LL term is topological so it is still enough to
consider the quadratic term and the solution for the PM parameters can
be obtained in a straightforward way. Up to a sign ambiguity in
$\beta_3$, all $\beta_n$ are uniquely fixed in terms of $l_2^g$.

In $d=7$, we also add the cubic Lovelock term and use the results of
section \ref{cubicll}. It turns out that there exists no nontrivial
solution to the constraint equations on the $\beta_n$. Hence there is
no PM theory in seven dimensions even with LL terms.

In $d=8$ it is still sufficient to consider only the quadratic and
cubic terms. Remarkably, the resulting PM candidate theory has two
free parameters. In particular, it is possible to set the cubic
coupling to zero and obtain a PM theory with only the quadratic
Lovelock term.

For $d=9,10$ it has been checked that adding both ${\cal L}_{(2)}$ and
${\cal L}_{(3)}$, or adding ${\cal L}_{(4)}$ alone, does not lead to
nontrivial solutions for the parameters. The only possibility then is
to add all terms up to quartic order, which we have not been able to
investigate with our methods at hand. Hence we can only make definite
statements about dimensions up to $d=8$. 

The results for the PM parameter values are summarized in Table
\ref{table}, where, to simplify the presentation, we define 
\be
\kappa^{(2)}_d\equiv \frac{m_g^{d-2}}{l^g_2m^d}\,,\quad
\kappa^{(3)}_d\equiv\frac{l^g_3m_g^{d-2}}{(l^g_2)^3m^d}\,.
\ee
\begin{table}[htdp]
\caption{Parameters of the PM candidates}
\begin{center}
\begin{tabular}{|c||c|c|c|c|c|c|}
\hline
dimension  & 3 & 4 & 5 & 6  & 8  \\
\hline
$l_2^g$ & -& -& $l_2^g$& $l_2^g$& $l_2^g$\\
\hline
$l_2^f$ & -& -& $\alpha^{-2} l_2^g$& $\alpha^{-2} l_2^g$&$\alpha^{-2} l_2^g$ \\
\hline
$l_3^g$ & -& -& -& -&  $l_3^g$ \\
\hline
$l_3^f$ & -& -& -&  -& $\alpha^{-4} l_3^g $ \\
\hline
$\beta_0$ &$\beta_0$ & $\beta_0$&$-\frac{5}{12}\kappa^{(2)}_5$ &$-\frac{5}{12}\kappa^{(2)}_6$  &$-\frac{21}{1600}\left(20\kappa^{(2)}_8+3\kappa^{(3)}_8\right)$    \\
\hline
$\beta_1$ & $\alpha\beta_0$&0 &$\frac{\alpha}{12}\kappa^{(2)}_5$ &0 &0   \\
\hline
$\beta_2$ &$\alpha^2\beta_0$ &$\frac{\alpha^2}{3}\beta_0$ & $-\frac{\alpha^2}{24}\kappa^{(2)}_5$&$\frac{\alpha^2}{12}\kappa^{(2)}_6$ &$-\frac{9\alpha^2}{1600}\kappa^{(3)}_8$  \\
\hline
$\beta_3$ & $\alpha^3\beta_0$& 0& $-\frac{\alpha^3}{24}\kappa^{(2)}_5$&$\pm\frac{\alpha^3}{6\sqrt{2}}\kappa^{(2)}_6$ &0  \\
\hline
$\beta_4$ & -&$\alpha^4\beta_0$ & $\frac{\alpha^4}{12}\kappa^{(2)}_5$&$\frac{\alpha^4}{12}\kappa^{(2)}_6$ &$\frac{3\alpha^4}{8000}\left(20\kappa^{(2)}_8-9\kappa^{(3)}_8\right)$  \\
\hline
$\beta_5$ & -& -& $-\frac{5\alpha^5}{12}\kappa^{(2)}_5$&0 &0  \\
\hline
$\beta_6$ & -& -& -&$-\frac{5\alpha^6}{12}\kappa^{(2)}_6$ &$-\frac{9\alpha^6}{1600}\kappa^{(3)}_8$   \\
\hline
$\beta_7$ & -& -& -& -&0   \\
\hline
$\beta_8$ & -& -& -& -&$-\frac{21\alpha^8}{1600}\left(20\kappa^{(2)}_8+3\kappa^{(3)}_8\right)$   \\
\hline
%$\beta_9$ & -& -& -& -& -& -& -& &  \\
%\hline
%$\beta_{10}$ & -& -& -& -& -& -& -& -&  \\
%\hline
\end{tabular}
\end{center}
\label{table}
\end{table}%

Note that in all cases the parameters  satisfy the symmetry property
(\ref{symans}) and the linear Higuchi bound,
\begin{align}
&d=5:\qquad m_{\mathrm{FP}}^2=-\frac1{4l^g_2}\left(1-\alpha c
+(\alpha c)^2\right)=\frac{\lambda}{2}\,,\nn\\
&d=6:\qquad m_{\mathrm{FP}}^2=-\frac1{3l^g_2}\left(1\pm\sqrt{2}\alpha c
+(\alpha c)^2\right)=\frac{2\lambda}{5}\,,\nn\\
&d=8:\qquad m_{\mathrm{FP}}^2=-\frac{3}{20l^g_2}\left(1
+(\alpha c)^2\right)=\frac{2\lambda}{7}\,.
\end{align}

\section{Discussion}
Our results are summarized in the introduction. One of the issues left
unanswered in the present analysis is the actual form of the gauge
transformation in the nonlinear PM theory. Currently one can see the
nonlinear form of this gauge transformation in the nonlinear theory in
dS backgrounds only for constant gauge parameters. Non-constant gauge
parameters will move the background away from dS. However, the
analysis of cubic interactions
\cite{Zinoviev:2006im,Joung:2012rv,Deser:2012qg} shows that, 
to cubic order, the theory is invariant under the full linear
transformation, and not just its constant part employed in this paper.
Another evidence for the existence of a nonlinear symmetry comes from
the study of cosmological solutions in \cite{vonStrauss:2011mq} based on
a FLRW-type ansatz. It turns out that for the PM values of parameters
(\ref{PMbs4}) and the cosmological ansatz of \cite{vonStrauss:2011mq},
the equations of motion leave a complete function of time undetermined
which indicates a nonlinear gauge invariance. Furthermore, using a
different approach, the authors in \cite{deRham:2012kf} consider
partial masslessness in massive gravity in a certain decoupling limit
and find a nonlinear symmetry that exists in that limit. This analysis
is not directly comparable to construction here, but could be taken as
another indication of the presence of the nonlinear symmetry.

\acknowledgments
We would like to thank E. Joung, M. Taronna and A. Waldron for 
discussions and comments. SFH would like to thank the organizers of
the Workshop on Supersymmetry, Quantum Gravity and Gauge Fields, 
September 12-14, 2012 in Pisa, Italy, where a preliminary version of 
these results were presented. 

\appendix

\section{Mathematical details}

Here we provide the details for deriving the equations of motion from
the bimetric action (\ref{Sgf}), computing the cosmological constants
on proportional backgrounds and finding the Fierz-Pauli mass of the
linear fluctuations. 
\subsection{Bimetric equations of motion}
The bimetric interaction potential in (\ref{Sgf}) is given in terms of 
$e_n(S)$ where $S=\sqrt{g^{-1}f}$. For a $d\times d$ matrix $S$, one
has  
\be
\det(\mathbb 1+\lambda S)=\sum_{n=0}^{d}\lambda^n\,e_n(S)\,.
\label{det-en}
\ee
$e_n(S)$ are the elementary symmetric polynomials of the eigenvalues of  
of $S$ and can be iteratively constructed starting with $e_0(S)=1$
and using the Newton's identities,  
\be
e_n(S)=-\frac{1}{n}\sum_{k=1}^{n}(-1)^k\,\tr(S^k)\,e_{n-k}(S)\,.
\label{it}
\ee
In $d$ dimensions, the iteration ends with $e_d(S)=\det S$ and then  
$e_n(S)=0$ for $n>d$. Obviously, $e_n(\lambda S)=\lambda^n e_n(S)$.
To obtain the equations of motion, one needs the variations, 
\be
\delta e_n(S)=-\sum_{m=1}^n(-1)^{m}\tr(S^{m-1}\delta S) \,e_{n-m}(S)\,.
\label{den}
\ee
These follow from (\ref{it}). Hence, 
\be
\delta[\det({S^{-1}})\, e_n(S)]=-\det({S^{-1}})\sum_{m=0}^n(-1)^{m}
\tr(S^{m-1}\delta S) \,e_{n-m}(S)\,.
\ee
If $S$ is the square-root of a matrix $E$, $S^2=E$, then 
\be
\tr(S^{m-1}\delta S)=\frac{1}{2}\tr({\sqrt E}^{m-2}\delta E) 
\ee
Note that for matrices, this relation holds only under the trace.
If $E=g^{-1}f$, then on varying $g^{-1}$, changing the summation
variable from $m$ to $r=n-m$, and cancelling $\sqrt{|f|}$, one gets,    
\be
\frac{\delta}{\delta g^{\mu\nu}}\left[\sqrt{|g|}
\,e_n(S)\right]=-\frac{1}{2}\sqrt{|g|}
(-1)^{n}\,g_{\mu\lambda}\,Y^\lambda_{(n)\nu}(S)\,,
\label{ddg-en}
\ee
where, the matrices $Y_{(n)\nu}^\mu(S)$ are given by,
\be
Y_{(n)}(S)=\sum_{k=0}^n (-1)^k \, S^{n-k} \,e_k(S).
\label{Yn}
\ee
Using (\ref{ddg-en}), for $V$ given by (\ref{V}) and satisfying the
property (\ref{f-g}), it is now straightforward to compute,
\begin{align}
&V_{\mu\nu}^g=\tfrac{1}{\sqrt{|g|}}\frac{\delta}{\delta g^{\mu\nu}}
\left[-2\sqrt{|g|} V\right]=g_{\mu\lambda}\,\sum_{n=0}^{d-1}(-1)^n
\beta_n\,Y_{(n)\nu}^\lambda(\sqrt{g^{-1}f})\,,
\label{Ap-Vmng}
\\
&V_{\mu\nu}^f=\tfrac{1}{\sqrt{|f|}}\frac{\delta}{\delta f^{\mu\nu}}
\left[-2\sqrt{|f|} V\right]=f_{\mu\lambda}\, \sum_{n=0}^{d-1}(-1)^n
\beta_{d-n}\,Y_{(n)\nu}^\lambda(\sqrt{f^{-1}g})\,.
\label{Ap-Vmnf}
\end{align}
This leads to the bimetric equations of motion (\ref{gf_eom}). Note
that the two equations are obtainable from each other by the
interchanges $\gmn \leftrightarrow \fmn$, $\beta_n 
\leftrightarrow \beta_{d-n}$, $m_g\leftrightarrow m_f$.  

\subsection{Proportional backgrounds and the cosmological constants}
Now consider the equations of motion (\ref{gf_eom}) for the background
ansatz $\bfmn=c^2\bgmn$, implying $\bar S=c\mathbb 1$. It is obvious
that the potential terms (\ref{Ap-Vmng}), (\ref{Ap-Vmnf}) become
cosmological contributions,   
\be
%\label{}
\tfrac{m^d}{m_g^{d-2}}\,\bar V_{\mu\nu}^{g}=\bgmn \Lambda_g\,,\qquad
\tfrac{m^d}{m_f^{d-2}}\,\bar V_{\mu\nu}^{f}=\bgmn \Lambda_f\,.
\label{VsLs}
\ee
$\Lambda_g$ and $\Lambda_f$ are expressed in terms of
$Y_{(n)}(c\mathbb 1)$. To compute this, consider (\ref{det-en}) for 
$\bar S=c\mathbb 1$. Since $\det(1+\lambda\bar S)=(1+\lambda c)^d$ and  
$e_k(c\mathbb 1)=c^k\,e_k(\mathbb 1)$, it follows that 
\be
e_k(\mathbb 1)={d\choose k}\equiv\frac{d !}{k !(d-k) !}\,,\qquad
\sum_{k=0}^n (-1)^k\, e_k(\mathbb 1)=
\sum_{k=0}^n (-1)^k\, {d\choose k}=(-1)^n{d-1\choose n}\,.
\label{ek1}
\ee
Exactly this sum appears in $Y_{(n)}(c\mathbb 1)$ (\ref{Yn}) which
becomes,
\be
Y_{(n)}(c\mathbb 1)=(-1)^nc^n{d-1\choose n}\,,\qquad
Y_{(n)}(c^{-1}\mathbb 1)=(-1)^nc^{-n}{d-1\choose n}\,.
\ee
On substituting in $\bar V_{\mu\nu}^{g,f}$ one reads off the
cosmological constants $\Lambda_g$ and $\Lambda_f$ from (\ref{VsLs})  
as,
\be
\Lambda_g=\frac{m^d}{m_g^{d-2}}\sum_{n=0}^{d-1}{ d-1\choose
  n}c^n\beta_n  \,,\qquad\quad
\Lambda_f=\frac{m^d}{m_f^{d-2}}\,c^{2-d}\sum_{n=1}^d{d-1\choose n-1}
c^n\beta_n\,.
\label{Ap-Ls}
\ee
Therefore, for proportional backgrounds, the bimetric equations reduce
to two copies of Einstein's equations (\ref{bg-g_eom}). The consistency of these
background equations with each other then requires
$\Lambda_g=\Lambda_f$, which generically determines $c$.

\subsection{Fluctuations and the Fierz-Pauli mass}
Let us now consider linear perturbations around the proportional
backgrounds (\ref{fcg}), 
\be
\gmn=\bgmn+\delta\gmn\,,\quad \fmn=\bfmn +\delta\fmn\,,
\label{fluc-A}
\ee
with $\bar f=c^2\bar g$. The corresponding linearized equations 
obtained from (\ref{gf_eom}) are,
\be
\label{l-gf_eom}
\delta\left(R_{\mu\nu}(g)-\tfrac{1}{2}\gmn R(g)\right)+
\tfrac{m^d}{m_g^{d-2}}\,\delta V_{\mu\nu}^{g}= 0\,,\qquad
\delta\left(R_{\mu\nu}(f)-\tfrac{1}{2}\fmn R(f)\right)+
\tfrac{m^d}{m_f^{d-2}}\,\delta V_{\mu\nu}^{f}= 0\,.
\ee
To compute these explicitly, we need, 
\begin{align}
&\delta V_{\mu\nu}^g\big\vert_{\bar S}=\frac{m_g^{d-2}}{m^d}
\Lambda_g\,\delta g_{\mu\nu}
+\bar g_{\mu\lambda}\,\sum_{n=1}^{d-1}(-1)^n
\beta_n\,\delta Y_{(n)\nu}^\lambda(S)\big\vert_{\bar S}\\
&\delta V_{\mu\nu}^f\big\vert_{\bar S}=
\frac{m_f^{d-2}}{m^d}\Lambda_f\,c^{-2}\delta f_{\mu\lambda}
+c^2{\bar g}_{\mu\lambda}\sum_{n=1}^{d-1}(-1)^n\beta_{d-n}\, 
\delta Y_{(n)\nu}^\lambda(S^{-1})\big\vert_{\bar S}\,,
\end{align}
The first term, obviously, is the contribution of the cosmological
constants to the linearized fluctuation equations. The second term
contains the FP mass and involve $\delta g$ and $\delta f$ only in the
combination $\delta S^\mu_{~\nu}=\frac{1}{2c}\bar g^{\mu\lambda}
(\delta f-c^2 \delta g)_{\lambda\nu}$. Hence the massive mode must
contain this combination of the fields. Since $\delta Y_0=0$, the
mass term is independent of both $\beta_0$ and $\beta_d$. From
(\ref{Yn}),
\be
\delta Y_{(n)}(S)\big\vert_{\bar S}=\sum_{r=0}^n(-1)^r\left[(n-r)c^{n-1}
\,e_r({\mathbb 1})\,\delta S + c^{n-r}\,\delta e_r\big\vert_{\bar S}
\right]\,.
\label{dYn}
\ee
The first term in $\delta Y_{(n)}\vert$, proportional to $\delta S$,
involves,  
\begin{align}
&n\sum_{r=0}^n(-1)^r\,e_r({\mathbb 1})=n\,(-1)^n {d-1\choose n}=
(d-1)\,(-1)^n {d-2\choose n-1}\\ 
&\sum_{r=1}^n(-1)^r r\,e_r({\mathbb 1})=d\,(-1)^{n}{d-2\choose n-1}  
\end{align}
To compute the second term in $\delta Y_{(n)}\vert$, proportional to
$\tr(\delta S)$, note that (\ref{den}) and (\ref{ek1}) give,
\be
\delta e_r\big\vert_{\bar S}=-c^{r-1}\,\tr(\delta S)\sum_{m=1}^r
(-1)^m\,e_{r-m}(\mathbb 1)=c^{r-1}\,\tr(\delta S)\,{d-1\choose r-1}\,.  
\ee
with $\delta e_0=0$. Then, the second term in (\ref{dYn}) contains,
using the second equation in (\ref{ek1}),
\be
c^{n-1}\,\tr(\delta S)\,\sum_{r=1}^n(-1)^r{d-1\choose r-1}=
c^{n-1}\,\tr(\delta S)\,(-1)^n {d-2\choose n-1}\,.
\ee
Putting these together in (\ref{dYn}) gives, for $n\geq 1$,
\be
\delta Y_{(n)}(S)\big\vert_{\bar S}=(-1)^n c^{n-1}{d-2\choose n-1}
\left[\tr(\delta S)\mathbb 1 - \delta S\right]\,.
\label{dYn2}
\ee
From this one can easily obtain $\delta Y_{(n)}(S^{-1})\big\vert_{\bar S}$ 
by replacing $c\rightarrow 1/c$ and noting that
$\delta(S^{-1})=-c^{-2}\delta S$. Finally, putting all this together
one gets,  
\begin{align}
&\delta V_{\mu\nu}^g\big\vert_{\bar S}=
\frac{m_g^{d-2}}{m^d}\Lambda_g\,\delta g_{\mu\nu}
+ N\,\bar g_{\mu\lambda}\,\left(\tr(\delta S)\delta^\lambda_\nu-\delta
  S^\lambda_{~\nu}\right)\,,
\label{dVmng}
\\
&\delta V_{\mu\nu}^f\big\vert_{\bar S}=
\frac{m_f^{d-2}}{m^d}\Lambda_f\,c^{-2}\delta f_{\mu\lambda}
-c^{2-d}\,N\, \bar g_{\mu\lambda}\,\left(\tr(\delta S)
\delta^\lambda_\nu-\delta S^\lambda_{~\nu}\right)\,,
\label{dVmnf}
\end{align}
with,
\be
N=\left[\sum_{n=1}^{d-1}\,c^{n-1}\beta_n{d-2\choose n-1}\right]\,.
\label{N}
\ee
The Einstein tensor in (\ref{gf_eom}) is linearized in the standard
way: one has, 
\be
\delta\left(R_{\mu\nu}+\frac{1}{2}\gmn R\right)=
\bar{\mathcal{E}}_{\mu\nu}^{\rho\sigma}\delta g_{\rho\sigma}+{\cal
  R}_{\mu\nu} \,. 
\label{dGg}
\ee
The structure of $\bar{\mathcal{E}}_{\mu\nu}^{\rho\sigma}
\delta g_{\rho\sigma}$ is given by (\ref{KO}) but with the background
metric $\bgmn$, and,
\be
{\cal R}_{\mu\nu}=-\frac{1}{2}\left(\bar R\delta\gmn-\bgmn
{\bar R}^{\rho\sigma}\delta g_{\rho\sigma}\right)
=-\frac{\Lambda_g}{d-2}\left(d\delta\gmn-\bgmn
{\bar g}^{\rho\sigma}\delta g_{\rho\sigma}\right)
\ee
In the last step we have used the background equation
(\ref{bg-g_eom}). Finally, using (\ref{dGg}) and the
corresponding equation for $\delta f$, along with (\ref{dVmng}) and   
(\ref{dVmnf}) in (\ref{l-gf_eom}), we arrive at the linearized
equations (\ref{l-g_eom}) and (\ref{l-f_eom}) for the fluctuations
$\delta g$ and $\delta f$.  

\section{The nonlinear $G$-$M^G$ action}
\label{App: G-M}
We sketch a brief derivation of the nonlinear bimetric action
\eqref{Sgf} written in terms of the fields 
\be\label{App: Gdef}
	G_{\mu\nu} = \left(1+(\alpha c)^{d-2}\right)^{(4-d)/(d-2)}
	\left(\gmn+(\alpha c)^{d-2}c^{-2}\fmn\right)\,,
\ee
and
\be\label{App: MdefG}
M^G_{\mu\nu}=\left(1+(\alpha c)^{d-2}\right)^{-2/(d-2)}
\left(G_{\mu\rho}S^\rho_{\ph\rho\nu}-c\,G_{\mu\nu}\right)\,,
\ee
where we recall that $S=\sqrt{g^{-1}f}$. These definitions are
consistent with all of our considerations in the main text. From these
definitions we straightforwardly obtain the inverted relations 
\be\label{App: gfGMrel}
\gmn = G_{\mu\rho}(\Phi^{-1})^\rho_{\ph\rho\nu}\,,\quad
\fmn =
G_{\mu\alpha}S^\alpha_{\ph\alpha\lambda}S^\lambda_{\ph\lambda\rho}
(\Phi^{-1})^\rho_{\ph\rho\nu}\,,
\ee
where, for notational purposes, we have defined
\be\label{App: Phidef}
\Phi^\rho_{\ph\rho\nu}=\left(1+(\alpha c)^{d-2}\right)^{(4-d)/(d-2)}
\left(\delta^\rho_\nu
+(\alpha c)^{d-2}c^{-2}S^\rho_{\ph\rho\sigma}S^\sigma_{\ph\sigma\nu}\right)\,.
\ee
From (\ref{App: MdefG}) we can also directly obtain $S$ in terms of
$\Gmn$ and $\Mmn^G$,  
\be\label{App: SGrel}
S^\rho_{\ph\rho\nu} = c\,\delta^\rho_\nu+\left(1+
(\alpha c)^{d-2}\right)^{2/(d-2)}G^{\rho\mu}\Mmn^G\,.
\ee
It is now easy to see that the volume densities can be rewritten
through 
\be\label{App: densRel}
\sqrt{g}=\sqrt{G}\det(\Phi)^{-1/2}\,,\quad
\sqrt{f}=\sqrt{G}\det(S)\det(\Phi)^{-1/2}\,.
\ee
In order to rewrite the interaction potential we note a general
relation obeyed by the symmetric polynomials, 
\be
e_n(1+\mathbb{X}) = \sum_{k=0}^n{d-k\choose n-k}e_k(\mathbb{X})\,.
\ee
If we consider the matrix parameterization $S =
c\left(\mathbb{1}+\mathbb{X}\right)$ (c.f. \eqref{App: SGrel}), we may
use this relation to rewrite the interaction potential, noting that 
\be
V(S;\beta_n)=\sum_{n=0}^d\beta_ne_n(S) = \sum_{n=0}^dc^n\alpha_n
e_n(\mathbb{X})	= V(\mathbb{X};c^n\alpha_n)\,,
\ee
where the different set of coefficients are related by
\be\label{App: alphabeta}
c^n\alpha_n=\sum_{k=n}^d{d-n\choose k-n}c^k\beta_k\,,\quad
c^n\beta_n=\sum_{k=n}^d(-1)^{n+k}{d-n\choose k-n}c^k\alpha_k\,.
\ee
This is straightforward to implement in the interaction potential,
which is then given by 
\be
 V(S;\beta_n) = V(G^{-1}M^G;\tilde\alpha_n)\,,\quad
\tilde\alpha_n=\left(1+(\alpha c)^{d-2}\right)^{2n/(d-2)}\alpha_n\,.
\ee
For the kinetic terms we choose to write these in terms of covariant
derivatives with respect to $\Gmn$. Using the Riemann curvature
definition
$[\nabla_\mu,\nabla_\nu]\,\omega_\rho=R_{\mu\nu\rho}^{\ph{\mu\nu\rho}\sigma}
\omega_\sigma$,
together with the general relation between two covariant derivations
on a manifold, 
\be
\nabla^g_\mu\omega_\nu=\nabla^G_\mu\omega_\nu-
C_{\mu\nu}^{\ph\mu\ph\nu\rho}\omega_\rho\,,\quad
C_{\mu\nu}^{\ph\mu\ph\nu\rho}=\tfrac{1}{2}g^{\rho\sigma}
\left(2{\nabla^G}_{(\mu}g_{\nu)\sigma}
-\nabla^G_\sigma \fmn\right)\,,
\ee
we obtain the relation (with an obvious similar relation for
$R_{\mu\nu}(f)$ obtained by replacing $g\rightarrow f$ in these
expressions) 
\be
R_{\mu\nu}(g)=R_{\mu\nu}(G)-2\nabla^G_{[\mu}C_{\rho]\nu}^{\ph{\rho]\nu}\rho}
+2C_{\nu[\mu}^{\ph{\nu[\mu}\sigma}C_{\rho]\sigma}^{\ph{\rho]\sigma}\rho}\,.
\ee
Tracing this with $g^{\mu\nu}$ ($f^{\mu\nu}$) and writing it out in
full we find that (modulo total derivatives and neglecting to write
out an overall factor of $\sqrt{|g|}$ ($\sqrt{|f|}$) on both sides) we
get Ricci curvature relations of the form 
\begin{align}\label{App: RgRGrel}
R(g) =&\, g^{\mu\nu}R_{\mu\nu}(G) 
-\tfrac1{2}g_{\rho\lambda}\nabla_\alpha g^{\rho\lambda}\nabla_\sigma 
g^{\alpha\sigma} +\tfrac1{2}g_{\rho\sigma}\nabla_\lambda 
g^{\alpha\sigma}\nabla_\alpha g^{\rho\lambda}\nn\\
&\,\,
-\tfrac1{4}g_{\rho\kappa}g_{\lambda\beta}g^{\alpha\sigma}\nabla_\alpha 
g^{\rho\lambda}\nabla_\sigma g^{\kappa\beta}
+\tfrac1{4}g_{\kappa\beta}g_{\rho\lambda}g^{\alpha\sigma}\nabla_\alpha 
g^{\kappa\beta}\nabla_\sigma g^{\rho\lambda}
\equiv g^{\mu\nu}R_{\mu\nu}(G) +\Pi^g\,,
\end{align}
with a similar relation for $R(f)$ and definition of $\Pi^f$, obtained
by replacing $g\rightarrow f$ in this. Collecting our results we can
now write the nonlinear action in terms of $\Gmn$ and $\Mmn^G$ as 
\begin{align}\label{App: SGM}
S_{GM}=m_g^{d-2}\int\td^dx\sqrt{|G|}&\det(\Phi)^{-1/2}
\Bigl[(\Phi G^{-1})^{\mu\nu}R_{\mu\nu}(G)
+\alpha^{d-2}\det(S)(\Phi G^{-1}S^{-2})^{\mu\nu}R_{\mu\nu}(G)\nn\\
&\qquad\quad+\Pi^g+\alpha^{d-2}\det(S)\,\Pi^f-2\tfrac{m^d}{m_g^{d-2}}\,
V(G^{-1}M^G;\tilde\alpha_n)\Bigr]\,,
\end{align}
where the relations (\ref{App: gfGMrel}), (\ref{App: Phidef}),
(\ref{App: SGrel}) and (\ref{App: RgRGrel}) can be used to get the
explicit form in terms of only $\Gmn$ and $\Mmn^G$. Expanding this action to second order 
gives the quadratic action \eqref{quadS} by construction.

\end{document}